\documentclass[aps,prd,onecolumn,showpacs,eqsecnum,nofootinbib]{revtex4}
\usepackage{epsfig}
\newcommand{\be}{\begin{equation}}
\newcommand{\ee}{\end{equation}}
\newcommand{\bea}{\begin{eqnarray}}
\newcommand{\eea}{\end{eqnarray}} 
\newcommand{\nn}{\nonumber}
\begin{document}     
\vskip-6pt \hfill {RM3-TH/08-3} \\     
\title{Solving the octant degeneracy with the Silver channel}

\author{Davide Meloni}
\email{meloni@fis.uniroma3.it}
\affiliation{        Dip. di Fisica, Universit{\`a} di Roma Tre and INFN, 
Sez. di Roma Tre,\\ Via della Vasca Navale 84, I-00146 Roma, Italy}

\begin{abstract}
We study the potential of the combination of the golden ($\nu_e \to \nu_\mu$) and 
silver ($\nu_e \to \nu_\tau$) channels 
to solve the octant degeneracy affecting the measurement of $\theta_{13}$ and
$\delta$ at future neutrino factories. To search for $\tau$ leptons produced in $\nu_\tau$
charged-current interactions, 
we consider two different detectors: the Emulsion Cloud Chamber 
detector (ECC) and the Liquid Argon Time Projection Chamber (LAr TPC).
We show that, when using similar detector masses, 
the upgraded version of the ECC detector 
(sensitive also to hadronic $\tau$ decay modes) and the LAr TPC detector have
comparable  sensitivities 
to the octant of $\theta_{23}$, being able to discriminate 
the two solutions for  $\sin^2 (2\theta_{13}) \gtrsim 10^{-3}$ 
at 3$\sigma$ level if $\theta_{23}=40^\circ$. We also show that the same 
setups are able to see deviation from maximal mixing as small as $\sim$ (4-6)\%
(at 3$\sigma$) if $\theta_{13}$ is close to its upper bound.

\end{abstract}



\maketitle

\section{Introduction}

The simultaneous measurement of the two still unknown neutrino mixing parameters 
$\theta_{13}$ and $\delta$ can be considered as one of the main 
neutrino-physics goal for the next decade \cite{ISS}. The ideal places where to look for the signal induced 
by the  small (vanishing ?) $\theta_{13}$ and the unconstrained phase $\delta$ are the appearance channels 
$\nu_e \to \nu_\mu$ \cite{Cervera:2000kp} and $\nu_e \to \nu_\tau$ \cite{Donini:2002rm}
(which then deserve the name of "golden" and "silver" channels, respectively) and any other 
transition obtained from them by symmetry relation like T or CP. The reason for this success 
is mainly ascribed to the leading $\theta_{13}$ dependence of the transition
probabilities and the next-to-leading dependence on $\delta$, a characteristic not shared by the
disappearance channels. If one uses the parametrization of the three-family leptonic mixing 
matrix $U_{PMNS}$ \cite{neutrino_osc} suggested in the PDG \cite{pdg}, $\theta_{13}$ and $\delta$ always
appear in the combination 
$\sin \theta_{13}\,e^{\pm i\,\delta}$ and a measurement of the CP-phase
$\delta$ cannot be achieved independently on $\theta_{13}$: the two parameters have to be measured
at the same time. Such a  correlation stains the achievable precision on the parameter
measurements since some $\theta_{13}$ and $\delta$  can reproduce
the same "true" physics (number of expected events) at some
confidence level, a situation which is known as the {\it
degeneracy problem}. The $(\theta_{13},\delta)$-pairs which mimic the value chosen by 
Nature are called {\it clone points} and come from our
ignorance about the "true" $\theta_{13}$ and $\delta$ (intrinsic clone), the sign of the
atmospheric mass difference (sign clones) and the octant to which $\theta_{23}$ belongs, if not
maximal (octant clones) \cite{Burguet-Castell:2001ez}-\cite{lin:2006}.
Many papers have already discussed the problem of the {\it eightfold} degeneracy and its possible 
solutions, the
common denominator being the need to combine as many independent informations as possible 
(spectral informations, different $L/E$ setups, different transition channels...) with the aim of 
locating the clone points in different region of the  $(\theta_{13},\delta)$ parameter space: in this way,  
the statistical significance of the fake solutions decreases and one can hope to solve the 
degeneracies \cite{Donini:2003vz}.

In this paper we want to
point out the importance of using the silver transition 
(in combination to the golden one) to solve the octant degeneracy at the neutrino factories
\cite{Geer:1997iz}. In the next section, we briefly revise some theoretical aspects related to
the location of the fake octant solutions in the ($\theta_{13}$,$\delta$)-plane and 
the promising synergy between the golden and silver channels; in Sect.~\ref{sect:star}, we discuss
the potential of an Emulsion Cloud Chamber 
detector (ECC) to help in solving the octant degeneracy and we discuss the better performance 
of an improved version of it, which allows to look for the hadronic $\tau$ decays, also 
(the {\it silver$^\star$} option, \cite{Huber:2006wb}); Sect.~\ref{sect:argon} is devoted to the study of the
octant degeneracy using the Argon Time Projection Chamber (LAr TPC) whereas in
Sect.~\ref{sect:maxmix} we briefly describe the possibility to establish
deviation from maximal mixing.
In Sect.~\ref{sect:concl} we eventually draw our conclusions. 

\section{The octant degeneracy}
\label{sect:octant}

The octant degeneracy originates because neutrino experiments are looking for $\nu_\mu, \nu_e$ disappearance
or $\nu_\mu \to \nu_\tau$ oscillations, where $\theta_{23}$ appears only through $\sin^2 2\theta_{23}$.
However, even if the dominant term in the $\nu_\mu \to \nu_\mu$ transition probability 
is completely symmetric under $\theta_{23} \to \pi/2 - \theta_{23}$, subleading effects could in
principle break this symmetry (although very difficult to isolate, the combination of appearance and
disappearance channels at 
a neutrino factory with L= 7000 Km can do the job for relatively large $\theta_{13}$, \cite{Donini:2005db}) and help in the
octant discrimination (see, e.g., \cite{Choubey:2005zy}-\cite{Hagiwara:2006nn}).
At the probability level, the octant degeneracy manifests itself in the $\nu_\alpha \to \nu_\beta$
transition  ($\alpha \neq \beta$) when the system of equations
\be
P^\pm_{\alpha \beta} (\bar \theta_{13}, \bar \delta,\bar \theta_{23}) = P^\pm_{\alpha \beta} (\theta_{13},
\delta,\pi/2-\theta_{23}) 
\label{eq:equi0}
\ee
admits solutions for $\theta_{13}$ and $\delta$ different from the values chosen by Nature (and
indicated with a bar). Here, $P^\pm_{\alpha \beta}$ are the transition probabilities for neutrinos and 
antineutrinos, respectively. The equations can be numerically solved in the
exact form; however, to get some light on the analytical behavior of the clone solutions, we expand the
probabilities $P^\pm_{\alpha \beta}$  in vacuum for
small $\theta_{13}$, $\Delta_\odot/\Delta_{atm}$  and
$\Delta_\odot L$  \cite{Cervera:2000kp}. We then obtain:
\bea
\label{eq:spagnoli}
\left \{ 
\begin{array}{lll}
P^\pm_{e \mu} (\bar \theta_{13}, \bar \delta) &=& 
X_\mu \sin^2 (2 \bar \theta_{13}) + 
Y \cos ( \bar \theta_{13} ) \sin (2 \bar \theta_{13} )
      \cos \left ( \pm \bar \delta - \frac{\Delta_{atm} L }{2} \right ) + Z_\mu   \\
 && \\
P^\pm_{e \tau} (\bar \theta_{13}, \bar \delta) &=& 
X_\tau \sin^2 (2 \bar \theta_{13}) -
Y \cos ( \bar \theta_{13} ) \sin (2 \bar \theta_{13} )
      \cos \left ( \pm \bar \delta - \frac{\Delta_{atm} L }{2} \right ) + Z_\tau 
\end{array}
\right .  
\eea

and
\be
\label{eq:emucoeff}
\left \{ 
\begin{array}{lll}
X_\mu &=& \sin^2 (\theta_{23} ) 
\sin^2 \left ( \frac{\Delta_{atm} L}{ 2 } \right ); \qquad 
X_\tau = \cos^2 (\theta_{23} ) 
\sin^2 \left ( \frac{\Delta_{atm} L}{ 2 } \right ) \ ;\\
\nn \\
Y &=& \sin ( 2 \theta_{12} ) \sin ( 2 \theta_{23} )
\left ( \frac{\Delta_\odot L}{ 2 } \right )
\sin \left ( \frac{ \Delta_{atm} L }{ 2 } \right ) \ ; \\
\nn \\
Z_\mu  &=& \cos^2 (\theta_{23} ) \sin^2 (2 \theta_{12})
\left ( \frac{\Delta_\odot L}{ 2 } \right )^2;
\qquad 
Z_\tau  = \sin^2 (\theta_{23} ) \sin^2 (2 \theta_{12})
\left ( \frac{\Delta_\odot L}{ 2 } \right )^2 \ .
\end{array}
\right .  
\ee 
with $\Delta_{atm} = \Delta m^2_{atm} / 2 E_\nu$ and $\Delta_\odot = \Delta m^2_\odot / 2 E_\nu$.
The three terms in eq.~(\ref{eq:spagnoli}) are called the {\it atmospheric}, {\it interference} and 
{\it solar} terms, respectively \cite{BurguetCastell:2002qx}. Two comments are in order.
The different sign in front of the interference term among the golden and silver transitions 
has been recognized as very useful to solve the intrinsic degeneracy for $\theta_{13}\gtrsim {\cal O }$ 
$(1^\circ)$ \cite{Donini:2002rm,Autiero:2003fu} (below that value the statistics of silver events
being too
small to be significant). The second important point is the different dependence on $\theta_{23}$ in 
$X_\mu, X_\tau$ and $Z_\mu, Z_\tau$ since $X_\tau = X_\mu ( \theta_{23} \to 
\pi/2 -\theta_{23})$ and $Z_\tau = Z_\mu ( \theta_{23} \to 
\pi/2 -\theta_{23})$, which is precisely the reason why we can profit of the silver channel to solve the
octant degeneracy.
In order to find simple analytical solutions of eq.~(\ref{eq:equi0}), we introduce the small parameter
$\varepsilon$, which represents the deviation of $\theta_{23}$ from maximal mixing, 
$\varepsilon = \theta_{23}-\pi/4$; then, a value $\varepsilon < 0$ ($\varepsilon > 0$) means that 
$\theta_{23}$ is in the first (second)
octant. Since it will be important for later applications, we remind that, in the atmospheric regime 
($X_{\mu,\tau} >> Z_{\mu,\tau}$) and
small $\varepsilon$, the intrinsic clones
are located in the fake points \cite{BurguetCastell:2002qx,Donini:2003vz}:

\be
\label{eq:int}
\left \{ 
\begin{array}{lll}
\delta_{\rm int} &\sim& \pi - \bar \delta 
\nn \\  &&  \\
(\sin^2 2\theta_{13})_{\rm int} &\sim& 
\sin^2 2\bar \theta_{13} \pm 4\, \cos \bar \delta\,\sin 2\theta_{12}\, 
\left ( \frac{\Delta_\odot L}{ 2 } \right ) \cot \left ( \frac{\Delta_{atm} L}{ 2 } \right ) + {\cal O}
(\varepsilon)
\end{array}
\right.   
\ee 
where the positive (negative) sign refers to the golden (silver) channel.
We recover the known result that the shift in $\theta_{13}$ is opposite for the two transitions
whereas the location in $\delta$ is quite independent on $\theta_{13}$.

Let us now go back to eq.~(\ref{eq:equi0}). We expect two different solutions: one of them is a 
mirror of the
true point ($S1$) whereas the other one is a mirror of the intrinsic clone in eq.~(\ref{eq:int}) ($S2$).
Working in the atmospheric regime (the one in which the silver statistics is
not too small compared to the golden one), and up to the first order in $\varepsilon$ 
(i.e., neglecting ${\cal O}$ ($\varepsilon \cdot \Delta_\odot \,L$) terms), we find:

\be\label{eq:sol1}
\hspace{-3cm} {\rm S1:~~~}
\left \{ 
\begin{array}{lll}
\sin \delta &\propto& (1 \mp 2\varepsilon)\,\sin \bar \delta \nn 
\\ && \\
\sin^2 2\theta_{13}  &\propto&  (1\pm 4 \varepsilon)\, \sin^2 2\bar\theta_{13} 
\end{array}
\right.
\ee
\be\label{eq:sol2}
\hspace{-1cm} {\rm S2:~~~}
\left \{ 
\begin{array}{lll}
\sin \delta &\propto& (\sin \delta)_{\rm int} \mp 2\varepsilon\,\sin \bar \delta \nn 
\\ && \\
\sin^2 2\theta_{13}  &\propto& (\sin^2 2\bar\theta_{13})_{\rm int} \pm 4 \varepsilon \,
\sin^2 2\bar\theta_{13}
\end{array}
\right.
\ee
where the upper (lower) sign refers to the golden (silver) channel and the subscripts $int$ refer to the
solution of eq.~(\ref{eq:int}).

It is quite interesting to observe that in $S1$ the location of the clones always goes in opposite
direction for golden and silver transitions, which means that the shifts $\theta_{13}-\bar \theta_{13}$ and 
$\delta-\bar \delta$ have different sign. Moreover, the distance between clones increases for 
larger deviations from maximal mixing. Notice that, in the limit $\varepsilon \to
0$, $S1$ reduces to the input point, which means that this solution is quite independent on $L/E$ and, then,
much more difficult to eliminate with methods based on energy resolution or different baselines. 

The solution $S2$ shows the interesting behavior that the octant clones are close to the 
intrinsic degeneracy; then, if the silver channel is useful to solve the intrinsic degeneracy, 
it could also be able to eliminate/reduce the impact of the octant degeneracy. 
Also in this case we can observe an opposite sign in front of the $\varepsilon$ term,
which slightly increases (reduces) the distance in $\theta_{13}$ between the golden and silver clones for 
$\varepsilon > 0$ ($\varepsilon < 0$). 

As a final remark, notice that the position in $\delta$ is 
{\it always} very close to the value of $\bar \delta$ (for $S1$) or $\delta_{\rm int}$ (for $S2$) since 
a shift of the sinus function proportional to $\varepsilon$ translates in a much smaller deviation of
the argument of the same function.

\begin{figure} [h!]
\centerline{\epsfig{figure=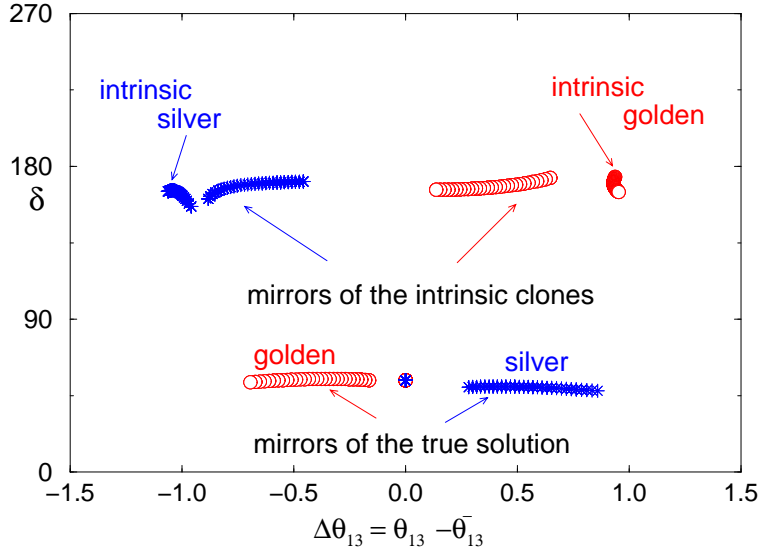,angle=0,width=10cm}}
\caption {\it Intrinsic and octant clone locations in the $(\Delta \theta_{13},\delta)$-plane. See text for
details. 
\label{fig:esempio}}
\end{figure}

The previous considerations, obtained from very simple analytical formulae 
in vacuum, describe quite well the general position of the octant clone points, even including matter 
effects, whose impact is neglible for $S1$ (due to the mild dependence on $L/E$) and more evident for $S2$
(but not so large to modify our conclusions, as we have carefully verified with numerical simulations). More
important, we can repeat the whole calculation using the number of expected events at a given
neutrino facility (instead of probabilities) and solving the analogous system of equations for rates in the exact form. 
In this way, the location of the clones will be much more similar to the experimental expectation since we
fold the probabilities with fluxes and cross sections \cite{Donini:2003vz}.
To be specific, we illustrate this point using two detectors able to measure the golden and 
silver transitions at the same baseline, $L=4000$ Km \footnote{In \cite{Huber:2006wb}, 
it has been shown that to maximize the statistical significance of the silver channel the best choice, and
also the most economic one, is to place the golden and silver detectors at the same baseline.}. Other
experimental details like absolute flux normalization, detector mass and overall efficiencies are unimportant since
they cancel when evaluating the integrated eq.(\ref{eq:equi0}).
The results obtained with this detector configuration and for $\bar \delta = 54^\circ$ and $\varepsilon =
-2^\circ$ are shown in Fig.~(\ref{fig:esempio}).

The points represent the location of the intrinsic and octant clones, 
for the golden and silver channels. For a fixed $\bar \theta_{13}$, we calculate the fake $\theta_{13}$
and $\delta$ and put them in the $(\Delta \theta_{13} = \theta_{13}- \bar \theta_{13}, \delta)$-plane. 
Varying the input $\bar \theta_{13}$, we are able to build the {\it clone flow} for a
given $\bar \delta$ \cite{Donini:2003vz}. All the features previously described can be clearly seen:
the golden and silver intrinsic clones are located in opposite directions ($\Delta \theta_{13} >0$ or 
$\Delta \theta_{13} < 0$), the mirrors of the intrinsic clones follow them (and are much
 less separated than
the intrinsic solutions thanks to the ${\cal O}$($\varepsilon$) terms in eq.~(\ref{eq:sol2})) whereas the
$\delta$ solutions of eq.~(\ref{eq:sol1}) show, as expected, a scarce dependence
on $\bar \theta_{13}$ and are very close to both the input values 
and $\delta_{int}$. From the theoretical point of 
view, all the clones are far away to each other and one can hope to solve the degeneracy. 

\section{Combining the MIND and ECC detectors}
\label{sect:star}

In order to verify how realistic these conclusions are, we 
perform a simulation of a neutrino factory in the spirit of \cite{Cervera:2000kp}, taking  
$1\cdot 10^{21}$ $\mu^+$ $\times$ 4 years, 
$1\cdot 10^{21}$ $\mu^-$ $\times$ 4 years \cite{ISS} and putting the
detectors at the same baseline $L=4000$ Km \footnote{Notice that the total number of stored muons, 
$8\cdot 10^{21}$, is in agreement
with the experimentally feasible $5\cdot 10^{20}$ muons per year and per baseline,
adopted  in the case of a neutrino factory pointing toward two detectors
located at two different baselines \cite{ISSreport}.}.
To look for golden muons, we use an improved version of the 
MIND detector described in \cite{dydak}: in particular, we
divide the signal and the background in $10$ energy bins of variable size \cite{anselmo}, according to the
energy resolution considered in \cite{anselmo2}. We also estimate efficiency and background from 
\cite{anselmo2}. To search for $\tau$ events we use the ECC detector, with efficiencies and backgrounds 
from \cite{Autiero:2003fu} and
divide the signal in 5 energy bins. The mass of the detector is 5 Kton. We also considered a 
conservative 10\% of systematic error for the silver detector and a 
2\% for the golden one.

The $\chi^2$ analysis, along the lines described in \cite{Donini:2002rm},
is performed simulating four input points, namely
$(\sin^2 2\bar \theta_{13},\bar \delta)=(1.2 \cdot 10^{-3},300^\circ),(2.7 \cdot 10^{-3},150^\circ),
(4.9 \cdot 10^{-3},0)$ and $(7.6 \cdot 10^{-3},90^\circ)$. 
The other oscillation parameters are fixed to the best fit values quoted in 
\cite{GonzalezGarcia:2007ib},
except for $\theta_{23}$, which we take away from maximal mixing, namely  $\theta_{23} = 43^\circ$ 
($\varepsilon = -2^\circ$). We used the GLoBES software \cite{Huber:2004ka} to compute the expected rates at the
neutrino factory.

\begin{figure} [h!]
\epsfig{figure=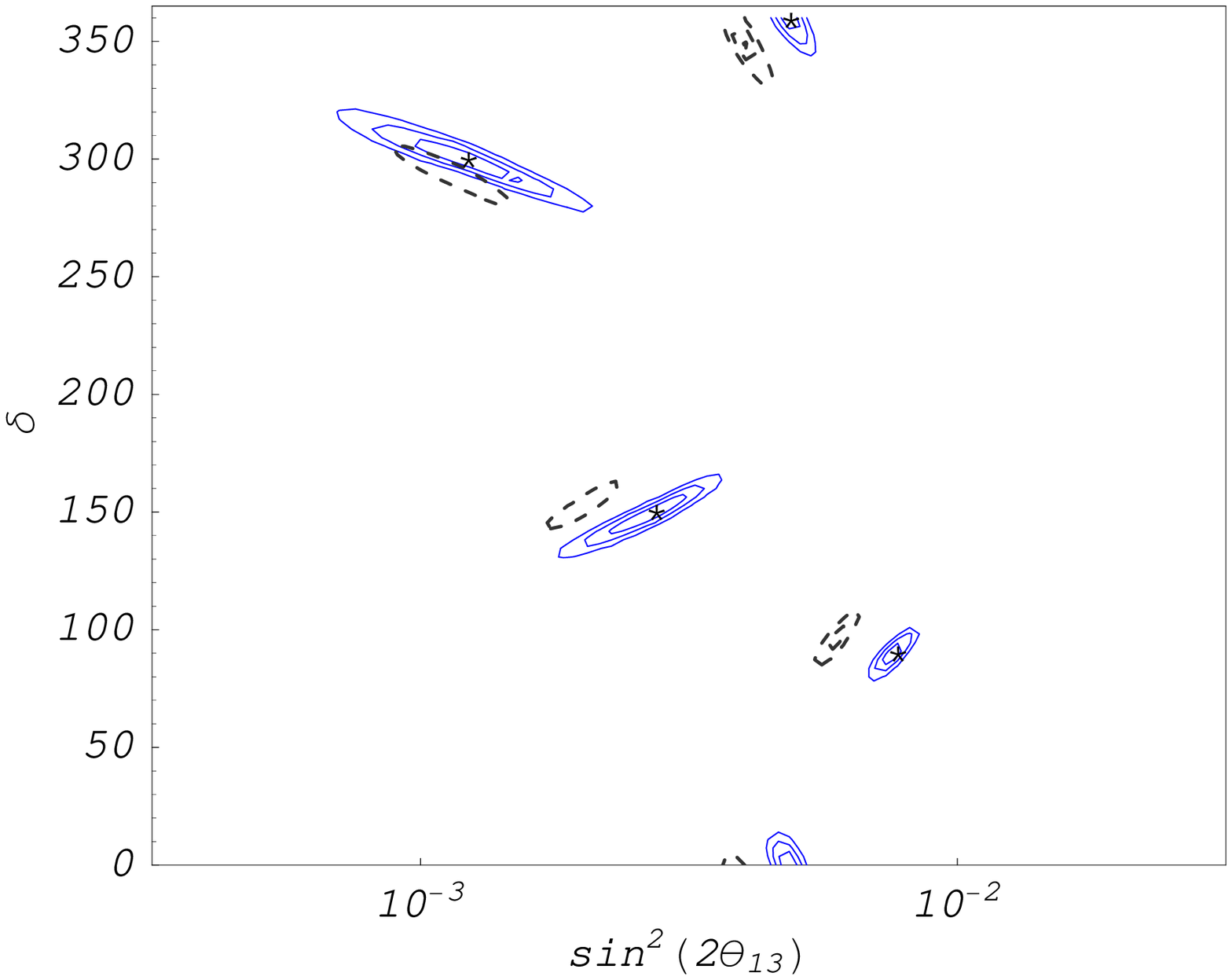,angle=0,width=8cm} \qquad 
\epsfig{figure=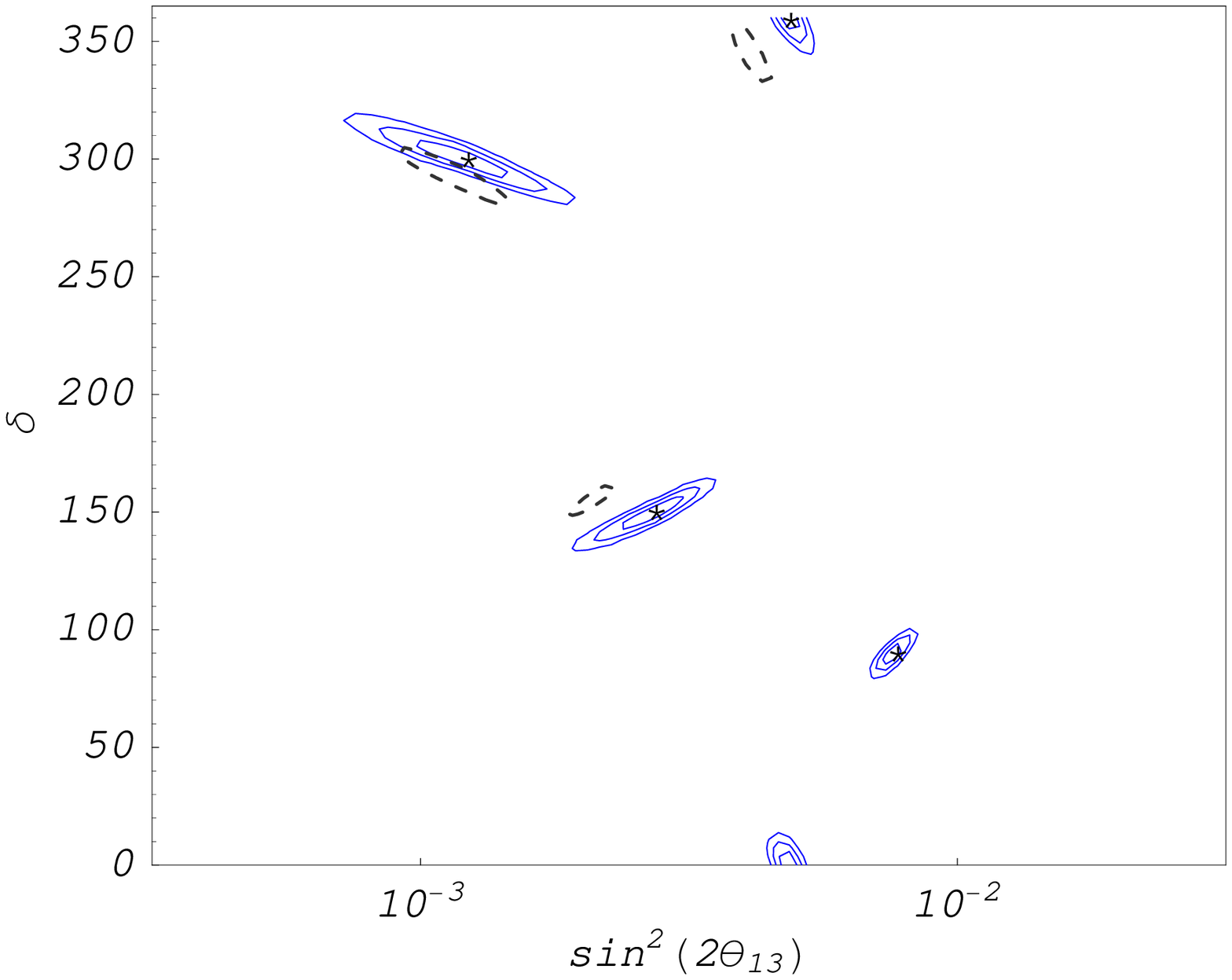,angle=0,width=8cm}
\caption {\it 1, 2 and 3$\sigma$ CL contours for the simulated points
$(\sin^2 2\bar \theta_{13},\bar \delta)=(1.2 \cdot 10^{-3},300^\circ),(2.7 \cdot 10^{-3},150^\circ),
(4.9 \cdot 10^{-3},0),(7.6 \cdot 10^{-3},90^\circ)$ in the 
$(\sin^2 2\theta_{13},\delta)$-plane. Continuum lines refer to the true and intrinsic points, dashed lines to the
octant clones. Left panel: results obtained combining the events from MIND and
ECC detectors. Right panel: results obtained using the combination 
MIND+ECC$^\star$ detectors.
\label{fig:due}}
\end{figure}

The impact of using the silver channel in solving the octant degeneracy (the intrinsic one being already 
solved for the representative points used in this analysis) can be seen in the left panel  
of Fig.~(\ref{fig:due}), in which we show the
1, 2 and 3$\sigma$ CL obtained combining the golden and silver events.
As it can be seen, the impact of the octant clones (dashed lines) 
in the measurement of $\theta_{13}$ and $\delta$ can be relevant; 
for any of the input points taken into account, the octant mirrors of the true points reduce the
achievable precision in the measurement of the unknowns, especially for $\theta_{13}$; moreover, 
according to our comments below 
eqs.~(\ref{eq:sol1}-\ref{eq:sol2}), the golden octant clones reside on the left of the input points whereas the value of
the fake $\delta$ is closed to that of the input and intrinsic points. 

We have chacked that very similar results can be obtained if we use the golden events only,  
mostly due to the quite good performance of the improved MIND detector and the relatively small 
statistics at the ECC detector.  It is clear that an improvement of
the silver detector is needed to take full advantage of the theoretical synergy with the golden channel 
espressed in eqs.~(\ref{eq:sol1}-\ref{eq:sol2}).

With the aim of improving the physics performance of the silver channel, some detector features have been already
discussed even though, at least to our knowledge, the results of a full simulation of the new version of the 
detector has not been published yet (see \cite{ISS}). Some new features rely on the possibility to performe a $dE/dx$ measurement to
reduce the charm background and to scan more than one brick per event, which would increases the signal detection
efficiency by about 20\% \cite{delellis}. Most notably, the possibility to search for the hadronic $\tau$ decay 
modes has been also discussed; it would allow a branching ratio gain ($\tau \to h$ and $\tau \to e$ channels) 
and a further reduction of
the background from hadronic interactions. This can be achieved embedding the ECC detector in a magnetic field and adding
a Total Active Scintillator Detector after the ECC in such a way to better discriminate between electrons and pions
\cite{delellis}
\footnote{It would also allow to look for the platinum transition $\nu_\mu \to \nu_e$.}.
These considerations inspired the authors of  \cite{Huber:2006wb} to introduce
the so-called $silver^\star$
(that, to avoid confusion with detectors, in the following will be called
ECC$^\star$). They assume
that the detector improvements allow an increase of the silver statistics by a factor of five and, thanks to the
improvements necessary for identifying hadronic tau decays, the background is  only a factor of three 
its standard value. In the following we will assume the ECC$^\star$ setup but we mantain the same detector mass used
before. With such a setup, the capability of the silver transition to solve the octant degeneracy is
improved, as we can see in the right panel of Fig.~(\ref{fig:due}), 
in which we plot the same contours as in the left panel 
but using the ECC$^\star$ setup.
For the larger $\sin^2 2\bar \theta_{13} = 7.6 \cdot 10^{-3}$, the clone point disappers whereas for the
other values of $\sin^2 2\bar \theta_{13}$ the clones only 
appear at 3$\sigma$ confidence level.

\begin{figure} [h!]
\epsfig{figure=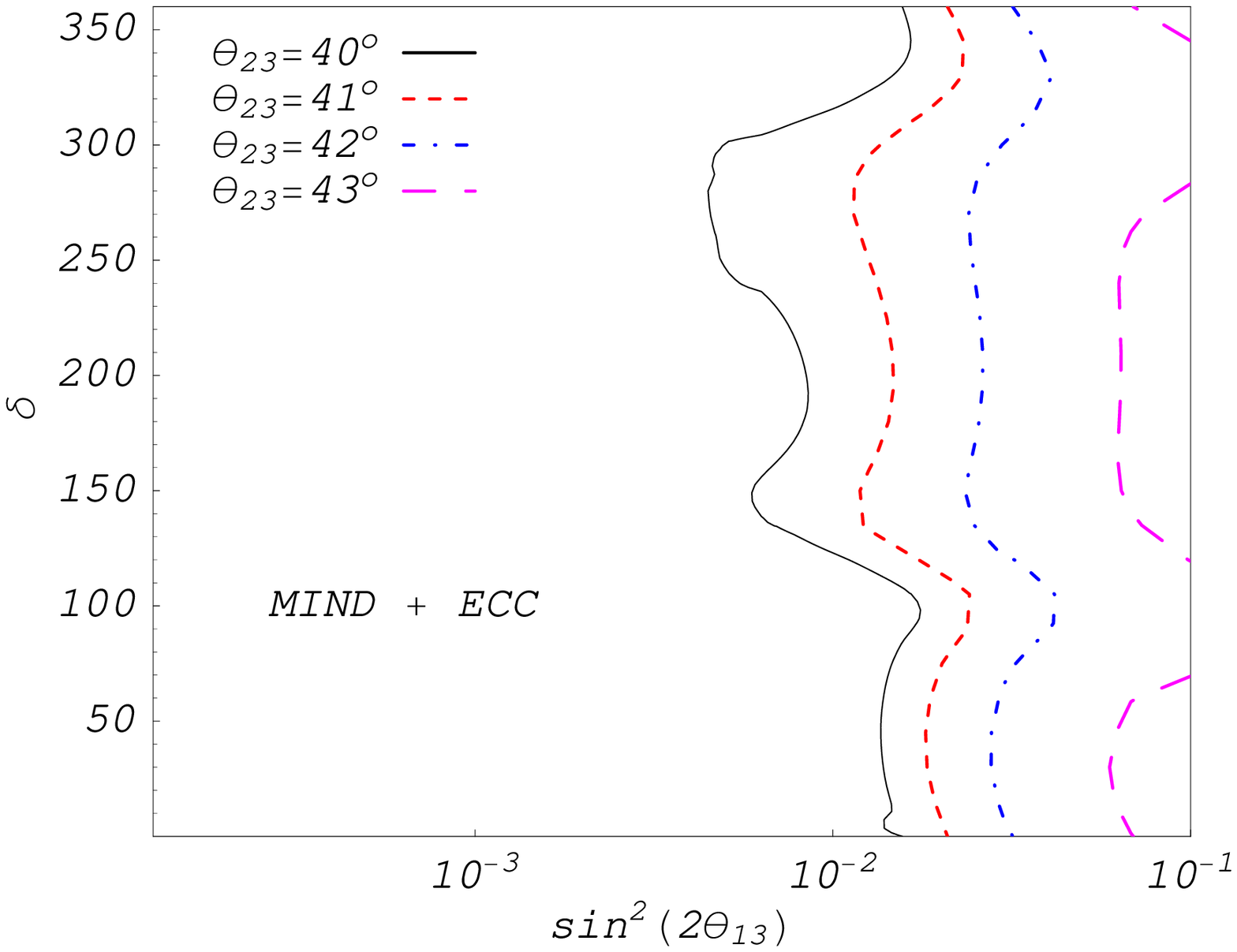,angle=0,width=8cm} \qquad 
\epsfig{figure=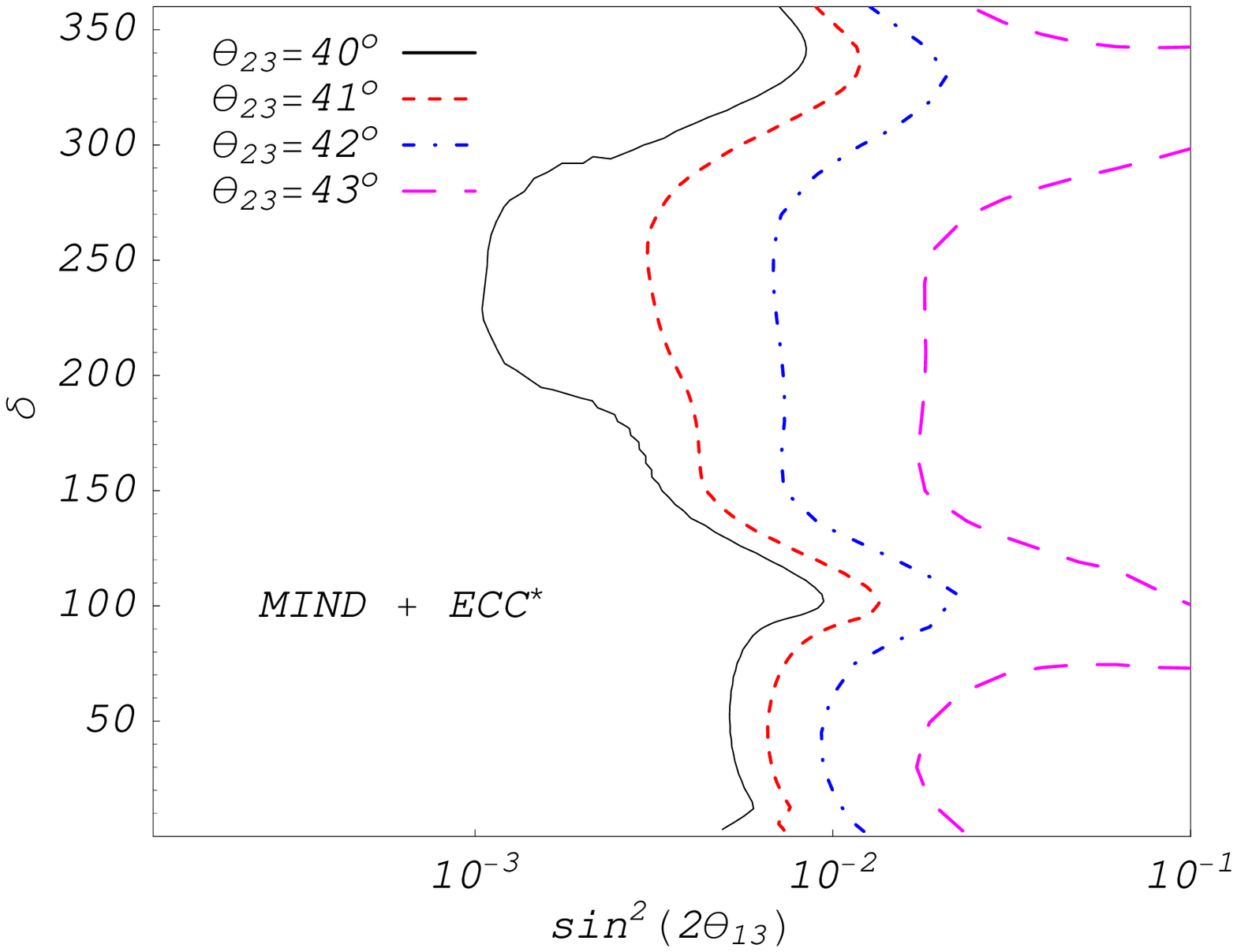,angle=0,width=8cm}
\caption {\it Sensitivity to the $\theta_{23}$ octant at 3$\sigma$ CL. 
The results are obtained combining the MIND+$ECC$ detectors for different $\theta_{23}$, 
from $40^\circ$ to $43^\circ$. In the left panel the sensitivity has been calculated 
using the standard ECC whereas in the right panel the ECC$^\star$ has been employed.
\label{fig:quattro}}
\end{figure}

In order to generalize our results, we compute the sensitivity to the $\theta_{23}$-octant: 
for any value of $\delta$, we look for the smaller value of $\theta_{13}$ for which no degenerate octant 
solutions appear at 3$\sigma$ CL. The results are shown in Fig.~(\ref{fig:quattro}).

In the left panel, we compute the sensitivity for the combination of MIND+standard ECC.
These results are simply explained with the help of 
eqs.~(\ref{eq:sol1}-\ref{eq:sol2}) since the distance between golden and silver octant clones is increased for larger
deviation from maximal mixing (that is, larger $|\varepsilon|$); thus, a much better performance is reached for 
$\theta_{23}=40^\circ$. 
The improvement reached using the ECC$^\star$ detector is clearly seen in the right panel; for each value of
$\theta_{23}$, a better performance by a factor of $2-3$ can be seen, depending on the value of the CP
phase considered.
In particular, in the range $\delta \in \left[200^\circ,300^\circ\right]$, the
sensitivity exhibits a sort of  plateau around 
$\sin^2(2\theta_{13})\sim 1.04 \cdot 10^{-3}$ for $\theta_{23}=40^\circ$ and 
$\sin^2(2\theta_{13})\sim 1.80 \cdot 10^{-2}$ for $\theta_{23}=43^\circ$,
whereas the minima are around $\delta \sim 100^\circ$. This is due to the fact
that for $\delta$ close to maximal CP violation, the number of golden 
events is extremely large and the silver statistics is comparably small in
such a way not to be able to help in solving the degeneracy\footnote{At
smaller confidence level, we have observed that the octant degeneracy can be 
solved also for $\sin^2(2\theta_{13})$ well below $10^{-3}$. This has to be
certainly ascribed to the action of the solar term $Z_\mu$ in the first
relation in eq.~(\ref{eq:spagnoli}), thus the golden channel alone is able to
get rid of the degeneracy for smaller $\theta_{13}$, for which the silver
statistics can be completely neglected.}.

\section{The Liquid Argon Time Projection Chamber}
\label{sect:argon}

The Liquid Argon Time Projection Chamber (LAr TPC) is a detector for uniform and high accuracy imaging of
massive active volumes. It is based on the fact that in highly pure Argon, ionization tracks can be drifted
over distances of the order of meters \cite{Aprili:2002wx}. 
Thanks to its high granularity target and its 
calorimetric properties, this design provides a clean identification and measurement of all three neutrino
flavours, and in particular $\nu_\tau$'s, to which we are interested in.
Following \cite{Bueno:2000fg}, we assume a non-magnetized target-detector and that charge discrimination
is only available for muons reaching and external magnetized-Fe spectrometer. Notice that 
a very large magnetized LAr TPC with a mass ranging from $\sim$ 10 to 100 Kton
has been recently proposed \cite{Rubbia:2004tz}. Our aim here is to estimate, depending on the adopted
detector mass, the potential of such a large detector in solving the octant degeneracy. As far as we know, a
dedicated simulation of the detector response to the silver channel has not been carried out, at least at the
sophisticated level reached in \cite{Autiero:2003fu}. This means that the estimate of  the background
affecting the measurement of the silver candidates events should rely to some {\it reasonable} assumption;
we then adopt the aptitude of considering all the background sources already described in
\cite{Autiero:2003fu} and to study the possible detector response to them \cite{campa}.
We consider $\tau$ lepton decays into muons, electrons and hadrons. The main background for decays into leptons $\tau \to
\ell \, \nu \nu$ comes from $\nu_\ell$ charged currents with charmed meson production.
To reduce these backgrounds to a tolerable level, we first observe that a cut in the muon momentum reduces the
background due to the decay of charged mesons at the level of $10^{-5}$ \cite{Bueno:2000fg}; we then apply the 
{\it loose kinematical cuts} described in \cite{Bueno:2000fg}, which allow to retain a good efficiency for 
the signal. This means a global efficiency for $\tau$ signal events of $\varepsilon_\tau = 0.334$ (including the branching
fractions to muons, electrons and hadronic channels) and a further reduction of the fractional background 
at the level of $5\cdot 10^{-8}$.

For hadronic decays, the most important source of background comes from neutral current events. Also in this case, a
fractional background at the level of $10^{-5}$ and a set of {\it loose kinematical cuts} reduce 
 the fractional background 
at the level of $2\cdot 10^{-7}$.
The other backgrounds for leptonic $\tau$ decays quoted in \cite{Autiero:2003fu} can be neglected.

\begin{figure} [h!]
\epsfig{figure=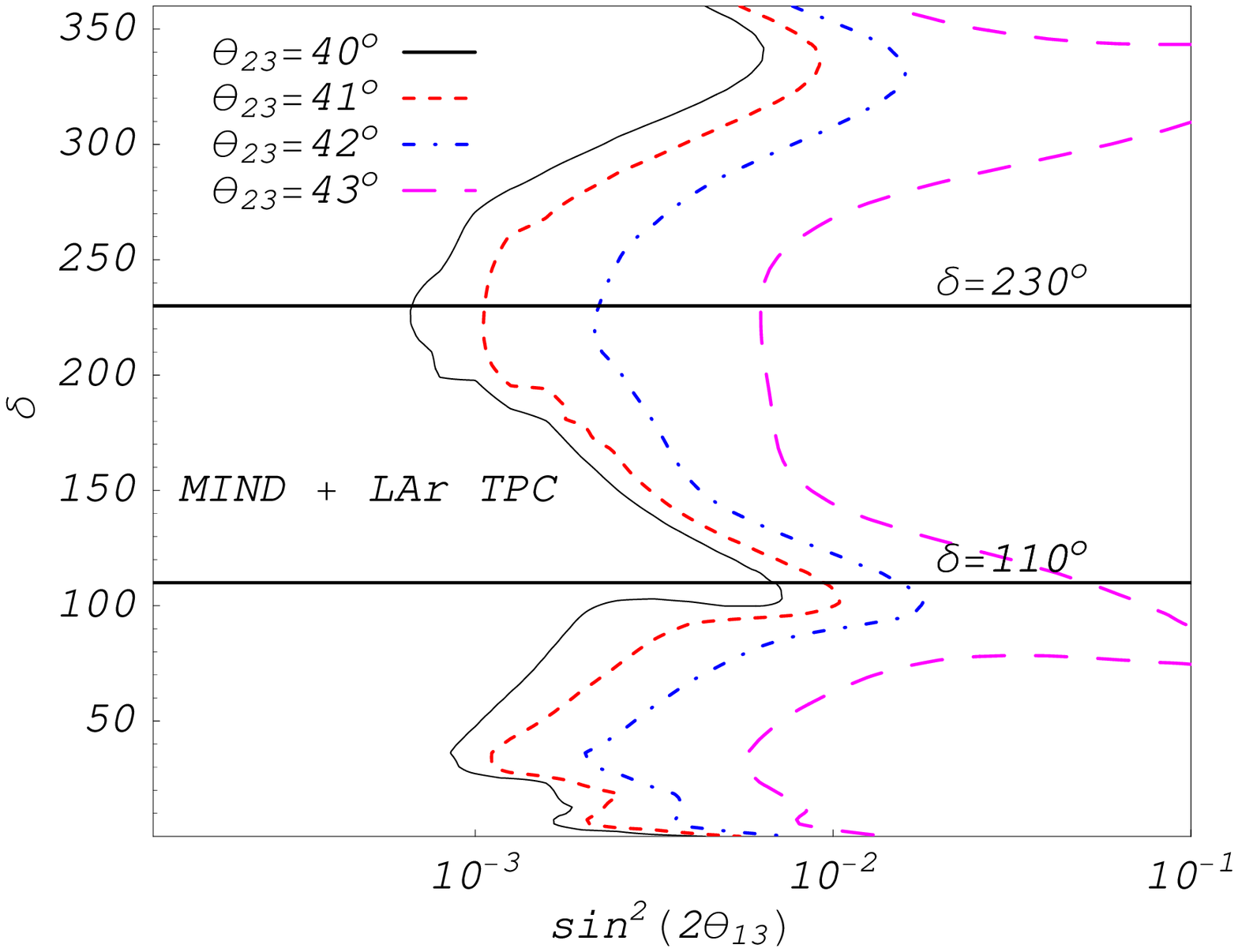,angle=0,width=8cm} \qquad 
\epsfig{figure=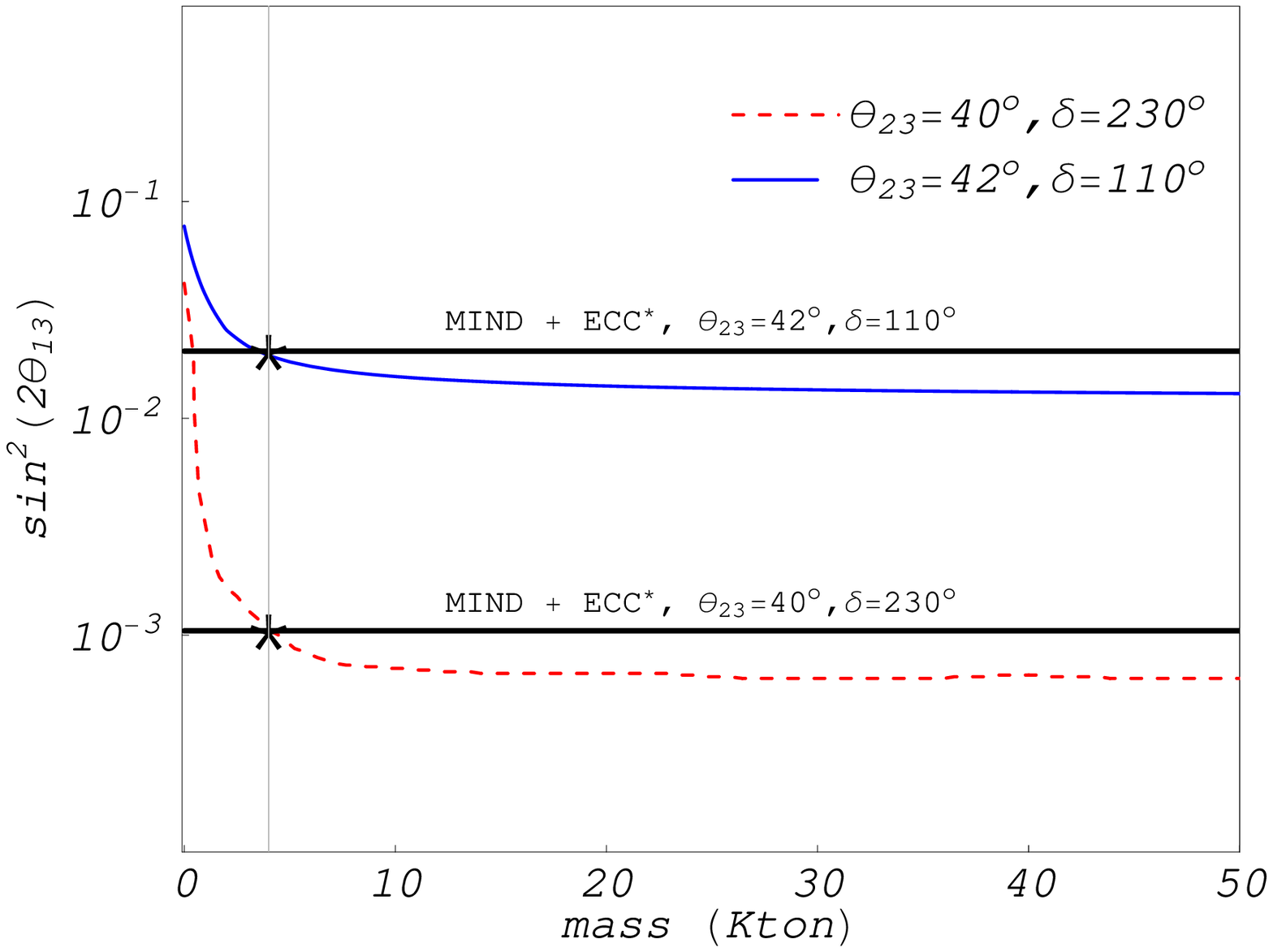,angle=0,width=8.2cm}
\caption {\it 
Left panel: Sensitivity to the $\theta_{23}$ octant at 3$\sigma$ CL  obtained combining the MIND and a 10 Kton LAr TPC 
detectors for different $\theta_{23}$, from $40^\circ$ to $43^\circ$.
The horizontal lines highlight the values $\delta=110^\circ$ and $230^\circ$.
Right panel: 3$\sigma$ CL sensitivity as a function of the LAr TPC detector mass. 
Continuum lines refer to $\theta_{23}=42^\circ$
whereas long-dashed lines are computed for $\theta_{23}=40^\circ$. The constant
horizontal lines represent the sensitivities of the  MIND+ECC$^\star$ combination at two
different points, $(\theta_{23},\delta)=(40^\circ,230^\circ)$
and $(42^\circ,110^\circ)$. See text for details.
\label{fig:icarus}}
\end{figure}

If we fix the detector mass to $10$ Kton (the minimum detector mass quoted for large LAr TPC project, see also 
\cite{ISS}), we observe a general improvement on the fit quality with respect to the results obtained with 
an ECC$^\star$. 
This reflects in a better sensitivity to the octant of $\theta_{23}$, as shown on the left panel of
Fig.~(\ref{fig:icarus}), to be compared with the right plot in Fig.~(\ref{fig:quattro}).

The 10 Kton LAr TPC  overwhelms the ECC$^\star$ by about a factor of 2-3 in sensitivity,  
the main reasons being the larger detector mass.
To verify this statement and compare both silver
detectors, we proceed in the following way. From Fig.~(\ref{fig:quattro}) and the left 
panel of Fig.~(\ref{fig:icarus}), we can see
that the sensitivity is maximal at $\delta \sim 230 ^\circ$ and minimal
at $\delta \sim 110 ^\circ$, almost independently of $\theta_{23}$. 
These values of $\delta$ are marked with horizontal lines in the left panel of 
Fig.~(\ref{fig:icarus}).
We then decide to performe the comparison using as a reference the sensitivity reach
of the MIND+ECC$^\star$ setup at the points $(\theta_{23},\delta)=(40^\circ,230^\circ)$
and $(42^\circ,110^\circ)$, in which $\sin^2(2\theta_{13})\sim 1.04 \cdot \,10^{-3}$ and 
$\sin^2(2\theta_{13})\sim 2.04 \cdot\,10^{-2}$, respectively. These values are shown in the
right panel of Fig.~(\ref{fig:icarus}) as continuum horizontal lines.
We can now compute the sensitivities of the MIND+LAr TPC setup at the same 
$(\theta_{23},\delta)$ points as
a function of the LAr TPC detector mass. The results are shown in the right panel 
of Fig.~(\ref{fig:icarus}), in which the continuum line refers to the worse case
$\theta_{23}=42^\circ$ and the long-dashed line to the best case 
$\theta_{23}=40^\circ$. The points in which these curves intersect the corresponding
MIND+ECC$^\star$ sensitivities define the values of the LAr TPC detector masses in which
the sensitivities to the $\theta_{23}$ octant are the same for both silver
detectors. It can be clearly seen that this happens for
masses just around 4 Kton (as stressed by the stars and 
the gray vertical line in the left panel of 
Fig.~(\ref{fig:icarus})), implying
a similar performance of both detectors when their masses are comparable. 
Notice that, as visible in the right panel of Fig.~(\ref{fig:icarus}), 
an increase in the LAr TPC detector 
mass above 15-20 Kton does not improve the sensitivity to the 
$\theta_{23}$ octant since  the systematic errors dominate over the statistical one and 
one cannot profit by the larger number of events.

\section{Deviation from maximal mixing}
\label{sect:maxmix} 
\begin{figure} [h!]
\centerline{\epsfig{figure=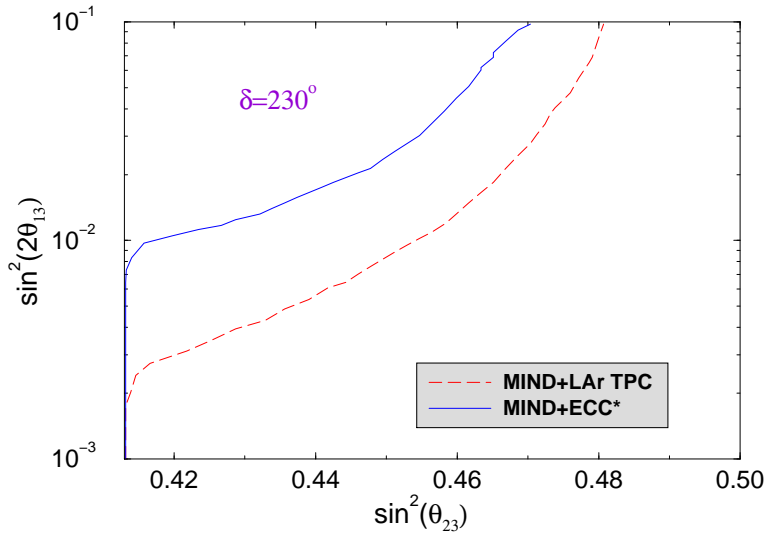,angle=0,width=10cm}}
\caption {\it 3$\sigma$ sensitivity to maximal $\theta_{23}$ computed for
$\delta=230^\circ$. Solid line refers to
the combination MIND+ECC$^\star$ whereas the long-dashed line refers to MIND+LAr TPC
setup. 
\label{fig:maxmix}}
\end{figure}

Having established the capability of the combination {\it golden+silver} to solve the octant ambiguity, 
we can now study the potential to exclude maximal $\theta_{23}$. We define the sensitivity of the deviation from
maximal mixing in the following way: for a given $\theta_{13}$, we look for the largest value of $\theta_{23}$ 
for which the two-parameter 3$\sigma$ contours do not touch $\theta_{23}=45^\circ$. The results also depend on the
assumed values of the atmospheric mass difference and the CP phase. 
While we retain $\Delta m^2_{atm}$ fixed to
its best fit quoted in \cite{GonzalezGarcia:2007ib}, we have analyzed the 
values $\delta=110^\circ$ and $\delta=230^\circ$, as in the previous section, and we
have found not a large difference. For this reason, in Fig.\ref{fig:maxmix} we only
quote the results obtained for $\delta=230^\circ$ in the 
$\left[\sin^2(\theta_{23})-\sin^2(2\theta_{13})\right]$-plane. The 
$\sin^2(\theta_{23})$ variable is used since we have considered both octants of
$\theta_{23}$ and we have found a behavior approximately symmetric of the
sensitivity. As we can see, the MIND+LAr TPC combination is more sensitive to the 
deviation from maximal mixing compared to the MIND+ECC$^\star$ result, an effect due to
the larger detector mass considered. For $\sin^2(2\theta_{13}) \sim 10^{-3}$, there
is fundamentally no possibility to tell a value for $\theta_{23}$ different from
maximal mixing (notice that the smaller $\sin^2(\theta_{23})$ corresponds to
$\theta_{23} \sim 40^\circ$). On the other hand, increasing $\theta_{13}$ one can
explore deviations from $\sin^2(\theta_{23}) = 0.5$: at the larger value of 
$\sin^2(2\theta_{13})$, corresponding approximately to the CHOOZ bound
$\theta_{13}\sim 13^\circ$ at 3$\sigma$ CL 
\cite{GonzalezGarcia:2007ib,Apollonio:2002gd}, deviation as small as 4\% (6\%) could be established
combining the information from MIND and LAr TPC detectors (MIND and ECC$^\star$). For comparison, 
we mention that the neutrino factory setup with baseline 
$L=7000$ Km described in \cite{Donini:2005db} is able to see deviations 
from maximal mixing at 3$\sigma$ for $\sin^2(\theta_{23})$ up to $\sim 0.47$ 
if $\sin^2(2\theta_{13}) = 7.6 \times 10^{-2}$ (and $\delta = 0$), a
value which is in between the results shown in  Fig.\ref{fig:maxmix}.

\section{Conclusions}
\label{sect:concl}
In this paper we have analyzed the potential of the combination of golden and silver transitions to solve the octant
degeneracy problem, affecting the future measurement of $\theta_{13}$ and $\delta$.
Starting form the transition probabilities in vacuum, we have shown that, in many cases, the octant clones are located 
in different points of the ($\theta_{13}$,$\delta$)-parameter space and that a promising synergy in solving this
degeneracy exists. Our naive results are confirmed with detailed numerical simulations for a standard 
neutrino factory, in which a MIND detector has been used to look for the golden transition and 
an improved  Emulsion Cloud Chamber detector and the Liquid Argon Time Projection Chamber 
have been employed to search for $\nu_e \to \nu_\tau$. As a general remark (almost 
independent on the silver detector), 
we observed that the octant degeneracy which mirrors the intrinsic clone gets solved in many of those cases in which the 
silver channel also eliminates the intrinsic degeneracy. However the octant clones which mirror the true points are much 
more difficult to eliminate, due to the very mild dependence on $L/E$ and the silver channel, mainly due to the low statistics, 
is not enough to get rid of them. This ultimately lowers the sensitivity to the octant of $\theta_{23}$.
It is obvious that the problem of statistics can be circumvented (or at least relaxed) using an improved ECC detector
(the ECC$^\star$ option described in the paper, which couple a magnetic field to the standard ECC) or massive liquid argon 
detector with masses ranging from 10 to 100 Kton.
A comparison between the physics potential of these two options strongly rely on the detector mass assumed to simulate the data.
The very conservative (and probably more realistic) hypothesis of using a 5 Kton mass for the ECC detector and 10 Kton for  
the LAr TPC shows that the latter would be better in sensitivity by roughly a factor of three 
whereas, if a similar 
detector mass is used (around 4-5 Kton), the sensitivity reach of the two detectors is comparable, with
 $\sin^2 (2\theta_{13}) \gtrsim 10^{-3}$ if $\theta_{23}=40^\circ$.

We have also analyzed the capability of the two options 
to exclude maximal mixing at 3$\sigma$ CL. We have found a
better performance of the 10 Kton LAr TPC detector, being able to see
deviations from $\theta_{23} = 45^\circ$ as small as 4\% if $\theta_{13}$ is 
close to its upper limit. Using the ECC$^\star$, the possibility to exclude
maximal mixing lowers to 6\%. 
 
We want to stress that these results have been obtained under reasonable assumptions for efficiency and
backgrounds of the LAr TPC detector and that a simulation dedicated to the silver channel is still missing. Notice also
that, contrary to what suggested in \cite{Huber:2006wb}, we did not increase
the mass of the ECC$^\star$, thus an
improvement of its performance is a possibility that needed to be investigated further.

\section{Acknowledgment}
D.M is strongly indebted to Andrea Donini for carefully reading the manuscript and for very useful
suggestions. D.M. wants also thank M. Campanelli and A. Cervera for very useful discussions.

\end{document}